\documentclass[%
 aip,
 jmp,%
 amsmath,amssymb,
 twocolumn,
]{revtex4}

\usepackage{graphicx}%
\usepackage{dcolumn}%
\usepackage{bm}%
\usepackage{color}
\usepackage{soul}
\usepackage{xcolor}

\begin{document}

\preprint{AIP/123-QED}

\title{A neural network potential with self-trained   atomic fingerprints:  \\ a test with the mW water potential}%

\def\correspondingauthor{\footnote{Corresponding author: john.russo@uniroma1.it}}

\author{Francesco Guidarelli Mattioli}
\author{Francesco Sciortino}
\author{John Russo\correspondingauthor{}}

\affiliation{ 
 Sapienza University of Rome, Piazzale Aldo Moro 2, 00185 Rome, Italy  
}%

\date{\today}%

\begin{abstract}

We present a neural network (NN) potential based on a new set of atomic fingerprints built upon two- and three-body contributions that probe distances and local orientational order respectively.   Compared to existing NN potentials, the atomic fingerprints depend on a 
small set of tuneable parameters  which are
 trained together with the neural network weights.  To tackle the simultaneous training of the 
 atomic fingerprint  parameters and neural network weights  we adopt  an  annealing protocol that progressively cycles 
 the learning rate, significantly improving the accuracy of the  NN potential. We test the  performance of the network potential
 against the mW model of water, which is a classical three-body potential that well captures the anomalies of the liquid  phase.
 Trained on just three state points, the NN potential is able to reproduce the mW model in a very wide range of densities and temperatures, from negative pressures to several $GPa$, capturing the transition from an open random tetrahedral network to a dense interpenetrated network. The NN potential also reproduces very well properties for which it was not explicitly trained, such as dynamical properties and the structure of the stable crystalline phases of mW.
\end{abstract}

\keywords{Suggested keywords}%
\maketitle

\section{INTRODUCTION}\label{sec:introduction}
Machine learning (ML) potentials represent one of the emerging trends in condensed matter physics and are  revolutionising the landscape of computational research. Nowadays, different methods to derive ML potentials have been proposed, providing a powerful methodology to model liquids and solid phases in a large variety of molecular systems~\cite{hansen2015machine,chmiela2017machine,haghighatlari2019advances,glielmo2020building,manzhos2020neural,gkeka2020machine,glick2020ap,benoit2020measuring,dijkstra2021predictive,unke2021machine,campos2021machine,musil2021physics,campos2022machine,goniakowski2022nonclassical,glielmo2022ranking,batzner20223,tallec2023potentials,batzner20223}. Among these methods, probably the most successful representation of a ML potential so far is given by Neural Network (NN) potentials, where the potential energy surface is the output of a feed-forward neural network ~\cite{behler2007generalized,behler2014representing,smith2017ani,smith2017ani2,schutt2017quantum,zhang2018end,singraber2019library,singraber2019parallel,husic2020coarse,cheng2020evidence,zhang2021phase,lu202186,tisi2021heat,behler2021four,zubatiuk2021development,jacobson2022transferable,gartner2022liquid,malosso2022viscosity}.

 In short, the idea underlying NN potentials construction is to train a neural network to represent the potential energy surface of a target system.
 The model is initially trained on a set of configurations generated ad-hoc, for which total energies and forces are known, by minimizing a suitable
 defined loss-function based on the error in the energy and force predictions. 
  If the training set is sufficiently broad and representative, the model can then be used to evaluate the total energy and forces of any related atomic configuration with an accuracy comparable to the original  potential.
Typically the original potential will include additional degrees of freedom, such as the electron density for DFT calculations, or solvent atoms in protein simulations, which make the full computation very expensive. By training the network only on a subset of the original degrees of freedom one obtaines a coarse-grained representation that can be simulated at a much reduced computational cost. NN potentials thus combine the best of two worlds,
retaining the accuracy of the underlying potential model,
  at the much lower cost of coarse-grained classical molecular dynamics simulations.
The accuracy of the NN potential depends crucially on how local atomic positions are encoded in the input of the neural network, which needs to retain the  symmetries of the underlying Hamiltonian, i.e. rotational, translational, and index permutation invariance. Several methods have been proposed in the literature~\cite{ko2021fourth,musil2021physics}, such as the approaches based on the Behler-Parrinello (BP) symmetry functions~\cite{behler2007generalized}, the Smooth Overlap of Atomic Positions (SOAP)~\cite{bartok2013representing}, N-body iterative contraction of equivariants (NICE)~\cite{nigam2020recursive} and polynomial symmetry functions
~\cite{bircher2021improved}, or frameworks like the DeepMD~\cite{zhang2018end}, SchNet~\cite{schutt2017quantum} and RuNNer~\cite{behler2007generalized}. In all cases, atomic positions are transformed into  atomic fingerprints (AFs). The choice of the AFs is particularly relevant, as it greatly affects the accuracy and generality of the resulting NN potential.

We  develop here a fully learnable NN potential in which the AFs, while retaining the simplicity of typical local fingerprints, do not need to be fixed beforehand but instead are learned during the training procedure.
The coupled training of the atomic fingerprint parameters  and of the network weights  makes  the  NN training process more efficient since the NN representation is spontaneously  built on a variable atomic fingerprint representation. 
To tackle the 
combined  minimization of the AF parameters  and of the network weights we adopt  an efficient  annealing procedure, that periodically  cycles the learning rate, i.e. the step size of the minimization algorithm,  resulting in a fast and accurate training process.

We validate the NN potential  on the mW model of water~\cite{molinero2009water}, which is a one-site classical potential that has found widespread adoption to study water's anomalies~\cite{russo2018water,holten2013nature} and crystallization phenomena~\cite{moore2011structural,davies2022accurate}.
Since the first pioneering MD simulations~\cite{barker1969structure,rahman1971molecular}, water
 is often chosen as a prototypical case study, as the large number of distinct local structures that are compatible with its tetrahedral coordination make it the molecule with the most complex thermodynamic behavior~\cite{tanaka2022roles}, for example displaying a liquid-liquid critical point at supercooled conditions~\cite{poole1992phase,cisneros2016modeling,debenedetti2020second,kim2020experimental,weis2022liquid}.
NN potentials for water have been developed starting from density functional calculations, with different levels of accuracy~\cite{nguyen2018comparison,cheng2019ab,gartner2020signatures,wohlfahrt2020ab,torres2021using,reinhardt2021quantum,lambros2021general,piaggi2022homogeneous}. NN potentials have also been proposed to parametrise  accurate classical models  for water with the aim of speeding up the calculations when  multi-body interactions are included~\cite{zhai2022short}, as in the MBpol model~\cite{babin2013development,babin2014development,medders2014development} or for testing the relevance 
 of the long range interactions, as for the SPC/E model~\cite{yue2021short}.
We choose the mW potential as our benchmark system because its explicit three-body potential term offers a challenge to the NN representation that is not found in molecular models built from pair-wise interactions.
We stress that we train the NN-potential against data which can be generated easily and for which structural and dynamic properties are well known (or can be evaluated with small numerical errors) in a wide range of temperatures and densities. In this way, we can perform a quantitative accurate comparison between the original mW model and the hereby proposed NN model.
 
Our results show that training the NN potential at even just one density-temperature state point provides an accurate  description of the mW model in a surrounding phase space region that is approximately a hundred kelvins wide.  A training based on three different state points extends  the convergence window extensively,
accurately reproducing  state points at extreme conditions, i.e. large negative and (crushingly) positive pressures.
We will show that the NN  reproduces thermodynamic, structural and dynamical properties of the mW liquid state, as well as structural properties of all the stable crystalline phases of mW water.

The paper is organized as follows. In Section \ref{sec:methods} we describe the new atomic fingerprints and the details about the Neural Network potential implementation, including the \emph{warm restart} procedure used to train the weights and the fingerprints at the same time. In Section~\ref{sec:results} we present the results, which include the accuracy of the models built from training sets that include one or three state points, and a comparison of the thermodynamic, structural and dynamic properties with those of the original mW model.
We conclude in Section~\ref{sec:conclusions}.

 \section{The Neural Network Model}\label{sec:methods}

\begin{figure*}[!t] %
   \centering
   \includegraphics[width=15.5cm]{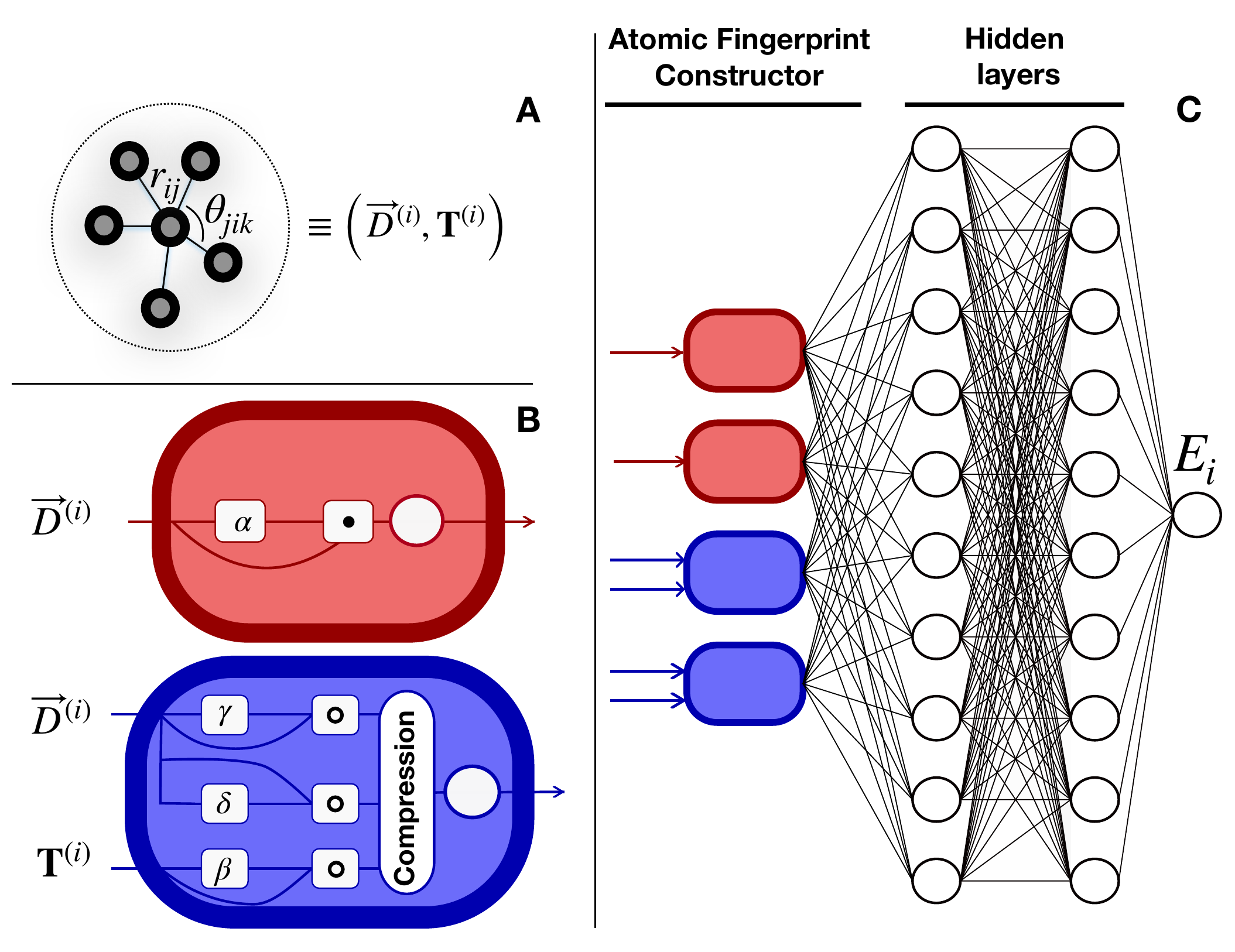} 
   \caption{Schematic representation of the Neural Network Potential flow.  (A) Starting from the relative distances and
   the triplets angles between neighbouring atoms, the input layer evaluates the atomic descriptors  $\vec{D}^{(i)}=\{D^{(i)}_{j}\}$ (Eq.~\ref{eq:internal1}) and $\bm{T}^{(i)}=\{T^{(i)}_{jk}\}$  (Eq.~\ref{eq:internal2}). (B) The first layer is the Atomic Fingerprint Constructor (AFC) and it combines the atomic descriptors into atomic fingerprints, weighting them with an exponential function. The red nodes perform the calculation of Eq.~\ref{eq:proj2bvec}, where from the two-body descriptors a weighting vector $\vec{D}_w^{(i)}(\alpha)=\{e^{\alpha D^{(i)}_j}\}$ is calculated (square with $\alpha$) and then the scalar product $\vec{D}^{(i)}\cdot\vec{D}_w^{(i)}(\alpha)$ is computed (square with point) and finally a logarithm is applied (circle). The blue nodes perform the calculation of Eq.~\ref{eq:proj3bvec}, where two weighting vectors are calculated from the two-body descriptors namely $\vec{D}_w^{(i)}(\gamma)$ and $\vec{D}_w^{(i)}(\delta)$ and one weighting matrix from the three-body descriptors $\bm{T}^{(i)}_w(\beta)=\{e^{\beta T^{(i)}_{jk}}/2\}$. Finally in the compression unit (Eq~\ref{eq:proj3bcomp}) values are combined as $0.5[\vec{D}^{(i)}\circ\vec{D}_w^{(i)}(\gamma)]^T[\bm{T}^{(i)}\circ\bm{T}^{(i)}_w(\beta)][\vec{D}^{(i)}\circ\vec{D}_w^{(i)}(\delta)]$ where we use the circle symbol for the element-wise multiplication. The output value of the compression unit is given to the logarithm function (circle). The complete network (D) is made of ten AFC units and two hidden layers with 25 nodes per layer and here is depicted 2.5 times smaller.}

   \label{fig:nn}
\end{figure*}

The most important step in the design of a feed-forward neural network potential is the choice on how to define the first and the last layers of the network, respectively named the \emph{input} and \emph{output} layers.
We start with the output layer, as it determines the NN potential architecture to be constructed. Here we follow the Behler Parrinello NN potential architecture~\cite{behler2007generalized}, in which the total energy of the system is decomposed as the sum of local fields ($E_i$), each one representing the contribution of a local environment centered around atom $i$. Being this a many-body contribution, it is important to note that $E_i$ is not the energy of the single atom $i$, but of all its environment (see also the Appendix A). With this choice, the total energy of the system is simply the sum over all atoms, $E=\sum E_i$, and the force $\vec f_i$ acting on atom $i$ is the negative gradient of the \emph{total} energy with respect to the coordinates $\nu$ of atom $i$, e.g. $f_{i\nu}=\partial E/\partial x_{i\nu}$. We have to point out that a NN potential is differentiable and hence it is possible to evaluate the gradient of the energy analytically. This allows to compute forces of the NN potential in the same way of other force fields, e.g. by the negative gradient of the total potential energy.

The input layer is built from two-body (distances) and three-body (angles) descriptors of the local environment, $\vec{D}^{(i)}$ and $\bm{T}^{(i)}$ respectively,  ensuring translational and rotational invariance.
The first layer of the neural network is the \emph{Atomic Fingerprint Constructor} (\emph{AFC}), as shown in Fig.~\ref{fig:nn}, which applies an exponential weighting on the atomic descriptors, restoring the invariance under permutations of atomic indexes. The outputs of this first layer are the atomic fingerprints (AFs) and in turn these are given to the first \emph{hidden layer}. We will show how this organization of the AFC layer allows for the internal parameters of the exponential weighting to be trained together with the weights in the hidden layers of the network. In the following we describe in detail the construction of the inputs and the calculation flow in the first layers.

\subsection{The atomic fingerprints}\label{sec:nn_model}

The choice of input layer presents considerably more freedom, and it is here that we deviate from previous NN potentials. The data in this layer should retain all the information needed to
properly evaluate forces and energies of the particles in the system, possibly exploiting the internal symmetries of the Hamiltonian (which in isotropic fluids are the rotational, translational and permutational invariance) to reduce the number of degenerate inputs.
Given  that the output was chosen as $E_i$, the energy of the atomic environment surrounding atom $i$, the input uses an atom-centered representation of the local environment of atom $i$.

In the input layer, we define an atom-centered representation of the local environment of atom $i$, considering both the distances $r_{ij}$ with the nearest neighbours $j$ within a spatial cut-off $R_c$, and the angles $\theta_{jik}$ between atom $i$ and the pair of neighbours $jk$ that are within a cut-off $R_c{'}$.
More precisely, for each atom $j$ within $R_c$ from $i$
we calculate the following descriptors

\begin{equation}
D^{(i)}_{j}(r_{ij};R_c)=
\begin{cases}
\frac{1}{2}\left[1+\cos\left(\pi\frac{r_{ij}}{R_c} \right)\right]\,\,\,\,\,\,r_{ij} \le R_c\\
0\,\,\,\,\,\,r_{ij}>R_c
\end{cases}
\label{eq:internal1}
\end{equation}

and, for each triplet $j-i-k$ within $R_c{'}$ from $i$,
\begin{eqnarray}\label{eq:internal2}
&&T^{(i)}_{jk}(r_{ij},r_{ik},\theta_{jik}) = \\ \nonumber
 &&\frac{1}{2}\left[1+\cos\left(\theta_{jik}\right)\right]~D^{(i)}_{j}(r_{ij};R_{c}{'})~D^{(i)}_{k}(r_{ik};R_{c}{'})
\end{eqnarray}

Here $i$ indicates the label of i-th particle while index $j$ and $k$ run over all other particles in the system.
In Eq.~\ref{eq:internal1}, $D^{(i)}_{j}(r_{ij};R_c)$ is a function that goes continuously to zero at the cut-off (including its derivatives). The choice of this functional form
guarantees that $D^{(i)}_{j}$ is able to express contributions even from neighbours close to the cut-off.  Other choices, based on polynomials or other non-linear functions, have been tested in the past~\cite{behler2021four}. For example, we tested a parabolic cutoff function which produced considerably worse results than the cutoff function in Eq.~\ref{eq:internal1}.
The function $T^{(i)}_{jk}(r_{ij},r_{ik},\theta_{jik})$ is also continuous at the triplet cutoff $R_c'$. The angular function
$\frac{1}{2}\left[1+\cos\left(\theta_{jik}\right)\right]$ guarantees that $0\le T^{(i)}_{jk}(r_{ij},r_{ik},\theta_{jik}) \le 1$. We note that the use of relative distances and angles in Eq.~\ref{eq:internal1}-\ref{eq:internal2} guarantees translational and rotational invariance.
  
The pairs and triplets descriptors are then fed to the \emph{AFC layer} to compute the atomic fingerprints, AFs.
These are computed by projecting the $D^{(i)}_j$ and $T^{(i)}_{jk}$  descriptors on a  exponential set of functions defined by

\begin{eqnarray}
{\overline{D}}^{(i)}(\alpha)&=&\ln \left [\sum_{j\neq i}D^{(i)}_j e^{\alpha D^{(i)}_j} +\epsilon \right ] -{Z_\alpha}\label{eq:proj2b}\\
{\overline T}^{(i)}(\beta,\gamma,\delta)&=&\ln \left [ \sum_{j\neq k\neq i}\frac{T^{(i)}_{jk}e^{\beta T^{(i)}_{jk}} e^{\gamma D^{(i)}_j} e^{\delta D^{(i)}_k}}{2}+\epsilon  \right ]\label{eq:proj3b} \\ 
&&- Z_{\beta\gamma\delta} \nonumber
\end{eqnarray}

These AFs are built summing over all pairs and all triplets involving particle $i$, making them invariant under permutations, and multiplying each descriptor by an exponential filter whose parameters are called $\alpha$ for distance AFs, and $\beta,~\gamma,~\delta$ for the triplet AFs.
These parameters play the role of feature selectors, i.e. by choosing an appropriate list of $\alpha,~\beta,~\gamma,~\delta$ the AFs can extract the necessary information from the atomic descriptors. The best choice of $\alpha,~\beta,~\gamma,~\delta$ will emerge automatically during the training stage.
In Eqs.~\ref{eq:proj2b}-\ref{eq:proj3b},
the number $\epsilon$ is set to $10^{-3}$ and fixes the value of energy in the rare event that no neighbors are found inside  the cutoff.
Parameters $Z_\alpha$ and $Z_{\beta\gamma\delta}$ are optimized during the training process, shifting the AFs towards positive or negative values,
and act as normalization factors that improve the representation of the NN.

The definitions in equations \ref{eq:proj2b}-\ref{eq:proj3b} can be reformulated in terms of product between vectors and matrices in the following way. The  descriptors in equations \ref{eq:internal1}-\ref{eq:internal2} for particle i can be represented as a vector  $\vec{D}^{(i)}=\{D_j^{(i)}\}$ and a matrix $\bm{T}^{(i)}=\{T_{jk}^{(i)}\}$ respectively. Given a choice of $\alpha,\,\beta,\,\gamma$ and $\delta$, three weighting vector $\vec{D}_w^{(i)}(\alpha)=\{e^{\alpha D^{(i)}_j}\}$, $\vec{D}_w^{(i)}(\gamma)=\{e^{\gamma D^{(i)}_j}\}$ and  $\vec{D}_w^{(i)}(\delta)=\{e^{\delta D^{(i)}_j}\}$ and one weighting matrix $\bm{T}^{(i)}_w(\beta)=\{e^{\beta T^{(i)}_{jk}}/2\}$ are calculated from $\vec{D}^{(i)}$ and $\bm{T}^{(i)}$.
The 2-body atomic fingerprint (Eq.~\ref{eq:proj2b}) is finally computed as  
\begin{equation}
\label{eq:proj2bvec}
{\overline{D}}^{(i)}(\alpha)=\ln \left[\vec{D}^{(i)}\cdot\vec{D}_w^{(i)}(\alpha)+\epsilon\right]-Z_{\alpha}
\end{equation}
The 3-body atomic fingerprint (Eq.~\ref{eq:proj3b}) is computed first by what we call \emph{compression} step  in Fig.~\ref{fig:nn} as 
\begin{equation}
\label{eq:proj3bcomp}
{\overline T_c}^{(i)}=\frac{[\vec{D}^{(i)}\circ\vec{D}_w^{(i)}(\gamma)]^T[\bm{T}^{(i)}\circ\bm{T}^{(i)}_w(\beta)][\vec{D}^{(i)}\circ\vec{D}_w^{(i)}(\delta)]}{2}
\end{equation}
and finally by 
\begin{equation}
\label{eq:proj3bvec}
{\overline T}^{(i)}(\beta,\gamma,\delta)=\ln \left[{\overline T_c}^{(i)}(\beta,\gamma,\delta)+\epsilon\right]-Z_{\beta\gamma\delta}
\end{equation}
where we use the circle symbol for the element-wise multiplication. The NN potential flow is depicted in Figure \ref{fig:nn} following the vectorial representation.\\
In summary, our AFs select the local descriptors useful for the reconstruction of the potential by weighting them with an exponential factor tuned with exponents $\alpha,\,\beta,\,\gamma,\,\delta$. A similar weighting procedure has been showed to be extremely powerful in the selection of complex patterns and is widely applied in the so-called \emph{attention layer} first introduced by Google Brain~\cite{vaswani2017attention}. However the AFC layer imposes additionally physically motivated constraints on the neural network representation.

We note that the expression for the system energy is a sum over the fields $E_i$, but the local fields $E_i$ are not additive energies, involving all the pair distances and triplets angles within the cut-off sphere centered on particle $i$.  This non-additive feature favours the NN ability to capture higher order correlations (multi-body contribution to the energy), and has been shown to outperform additive models in complex datasets~\cite{pozdnyakov2020incompleteness}. The NN non-additivity requires the derivative of the whole energy $E$ (as opposed to $E_i$) to estimate the force on a particle $i$. In this way, contributions to the force on particle $i$ come not only from the descriptors of $i$ but also from the descriptors of all particles who have $i$  as a neighbour, de facto enlarging the effective region in space where
interaction between particles are included. This allows the network to include contributions from length-scales larger than the cutoffs that define the atomic descriptors. The Appendix A provides further information on this point.

\subsection{Hidden layers}

We employ a standard feed-forward fully-connected neural network composed of two hidden layers with 25 nodes per layer and using the hyperbolic tangent ($\tanh$) as the activation function.
The nodes of the first hidden layer are fully connected to the ones in the second layer, and these connections have associated weights $W$ which are optimized during the training stage.

The input of the first hidden layer is given by the AFC layer where we used five nodes for the two-body AFs (Eq.~\ref{eq:proj2b}) and five nodes for the three-body AFs (Eq.~\ref{eq:proj3b}) for a total of 10  AFs for each atom. We explore the performance of some combinations for the number of two-body and three-body AF in Appendix D and we find that the choice of five and five is the more efficient.

The output is the local field $E_i$, for each atomic environment $i$, whose sum $E=\sum_{i=1}^NE_{i}$ represents the NN estimate of the potential energy $E$  of the whole system.

\subsection{Loss function and training strategy}\label{sec:nn_implementation}

To train the NN-potential we minimize a loss function computed over $n_f$ frames, i.e. the number of independent configurations extracted from an equilibrium simulation of the liquid phase of the target potential (in our case the mW potential). The loss function is the sum of two contributions.

The first contribution, $H[\{\Delta \epsilon^k, \Delta f_{i\nu}^k\}]$, expresses the difference in each frame $k$ between the NN estimates and the target values for both the total potential energy (normalized by total number of atoms) $\epsilon^k$ and the atomic forces $f_{i\nu}^k$ acting in direction $\nu$ on  atom $i$.
The $n_f$ energy $\epsilon^k$  values  and  $3 N n_f$  force $f_{i\nu}^k$  values are combined in the following expression
 
 \begin{eqnarray}
 H[\{\Delta \epsilon^k, \Delta f_{i\nu}^k\}]=\frac{p_e}{n_f}\sum_{k=1}^{n_f}h_{\text{Huber}}(\Delta \epsilon^k)  + \nonumber \\
 \frac{ p_f }{3N n_f}\sum_{k=1}^{n_f}\sum_{i=1}^{N} \sum_{\nu=1}^3h_{\text{Huber}}(\Delta f_{i\nu}^k) 
 \label{eq:defhubervec}
 \end{eqnarray}
 where $p_e=0.1$ and $p_f=1$ control the relative contribution of the energy and the forces to the loss function, and $h_\text{Huber}(x)$ is the so-called Huber function
 \begin{eqnarray}
h_\text{Huber}(x)=
\begin{cases}
0.5x^2\,\,\,\,{\rm if} \,\,|x| \le 1\\
0.5+(|x|-1)\,\,\,\,{\rm if} \,\,|x|>1\\
\end{cases}\label{eq:defhuberscal}
\end{eqnarray}
$p_e$ and $p_f$ are hyper-parameters of the model, and we selected them with some preliminary tests that found those values to be near the optimal ones.
The Huber function~\cite{10.1214/aoms/1177703732} is an optimal choice whenever the exploration of the loss function goes through large errors caused by outliers, i.e. data points that differ significantly from previous inputs.
Indeed when a large deviation between the model and data occur, a mean square error minimization may gives rise to an anomalous trajectory in  parameters space, largely affecting the stability of the training procedure. This may happen especially in the first part of the training procedure when the parameter optimization, relaxing both on the  energy and forces error  surfaces  may experience some instabilities. 

The second contribution to the loss function is 
a regularization function, $R[\{\alpha^l,\beta^m,\gamma^m,\delta^m\}]$, that serves to limit the
range of positive values of $\alpha^l$ and of the triplets $\beta^m,\gamma^m,\delta^m$ 
(where the indexes $l$ and $m$ run over the five different values of $\alpha$ and five different triplets of values for $\beta$, $\gamma$ and $\delta$) in the window $-\infty$ to $5$.  To this aim we select the
commonly used  relu  function 
\begin{eqnarray}
r_\text{relu}(x)=
\begin{cases}
x-5\,\,\,\,\,{\rm if}\,\,x>5\\
0\,\,\,\,\,{\rm if}\,\,x\le 5
\end{cases}\\
\end{eqnarray}
 
and write
 
\begin{eqnarray}
R[ \{ \alpha^l, \beta^m,\gamma^m,\delta^m  \}  ]=\sum_{l=1}^{5}  r_\text{relu}(\alpha^l) + \nonumber \\
\sum_{m=1}^{5}  [r_\text{relu}(\beta^m)+r_\text{relu}(\gamma^m) +r_\text{relu}(\delta^m) ]
\label{eq:relu}
\end{eqnarray}

Thus, the $R$  function is activated whenever one parameters of the \emph{AFC layer} becomes, during the minimization, larger  than 5.

To summarize, the global loss function $\mathcal{L}$  used in the training of the NN is 
\begin{eqnarray}
\mathcal{L}[\epsilon,f]=H[\{\Delta \epsilon^k, \Delta f_{i\nu}^k\}]  + p_b R[ \{ \alpha^l, \beta^m,\gamma^m,\delta^m  \}  ]
\label{eq:huber}
\end{eqnarray}
where $p_b=1$ weights the relative contribution of $R$ compared to $H$. 

\vskip 0.3cm

Compared to a standard  NN-potential, we train not only   
the network weights $W$ but also the AFs parameters $\Sigma \equiv \{ \alpha^l, \beta^m,\gamma^m,\delta^m  \}$ at the same time. The simultaneous  optimization  of the weights W and AFs $\Sigma$ prevents possible  bottleneck in the optimisation of W at fixed representation of $\Sigma$. Other  NN potential approaches 
implement a separate initial procedure to optimise the $\Sigma$ parameters followed by the optimisation of W at   fixed  $\Sigma$ \cite{imbalzano2018automatic}. The two-step procedure not only requires a specific methodological choice for optimising $\Sigma$, but also
 may not  result in   the optimal values, compared to a search in the full parameter space (i.e.  both $\Sigma$ and W).
Since the complexity of the loss function has increased, we have
investigated in some detail some efficient strategies that lead to a fast and accurate training.
Firstly, we initialize the parameters  $W$ via the Xavier algorithm, in which the weights are extracted from a random uniform distribution~\cite{glorot2010understanding}. To initialize the $\Sigma$  parameters we used a uniform distribution in interval $\left[-5,5\right]$.
We then minimize the loss function using the \emph{warm restart procedure} proposed  in reference ~\cite{loshchilov2016sgdr}.
In this procedure, the learning rate $\eta$ is reinitialized at every cycle $l$  and inside each cycle it decays as a function of the number of training steps $t$ following

\begin{eqnarray}
\eta^{(l)}(t)= A_l\left \{\frac{(1-\xi_f)}{2}    \left[1+\cos\left(\frac{\pi t} {T_l}\right)\right]+\xi_f\right \} && \label{eq:lr} \\
 0 \le t \le T_l && \nonumber
\end{eqnarray}
where  $\xi_f=10^{-7}$, 
$ A_l=   \eta_0\xi_0^l $ is the initial learning rate of the $l$-th cycle with $\eta_0=0.01$ and $\xi_0=0.9$, $T_l= b \tau^l$ is the period  of the $l$-th cycle with $\tau=1.4$ and $b=40$.
The absolute number of training steps  $n$ during cycle $l$  can be calculated summing over the length of all previous cycles as  $n=\tau+ \sum_{m=0}^{l-1} T_m $.

We also select to evaluate the loss function for groups of four frames (mini-batch) and we randomly select $200$ frames $n_f=200$ for a system of 1000 atoms and hence we split this dataset in $160$ frames $(\%80)$ for the training set and the  $40$ frames $(\%20)$ for the test set.

\begin{figure}[!t] %
   \centering
   \includegraphics[width=8.5cm]{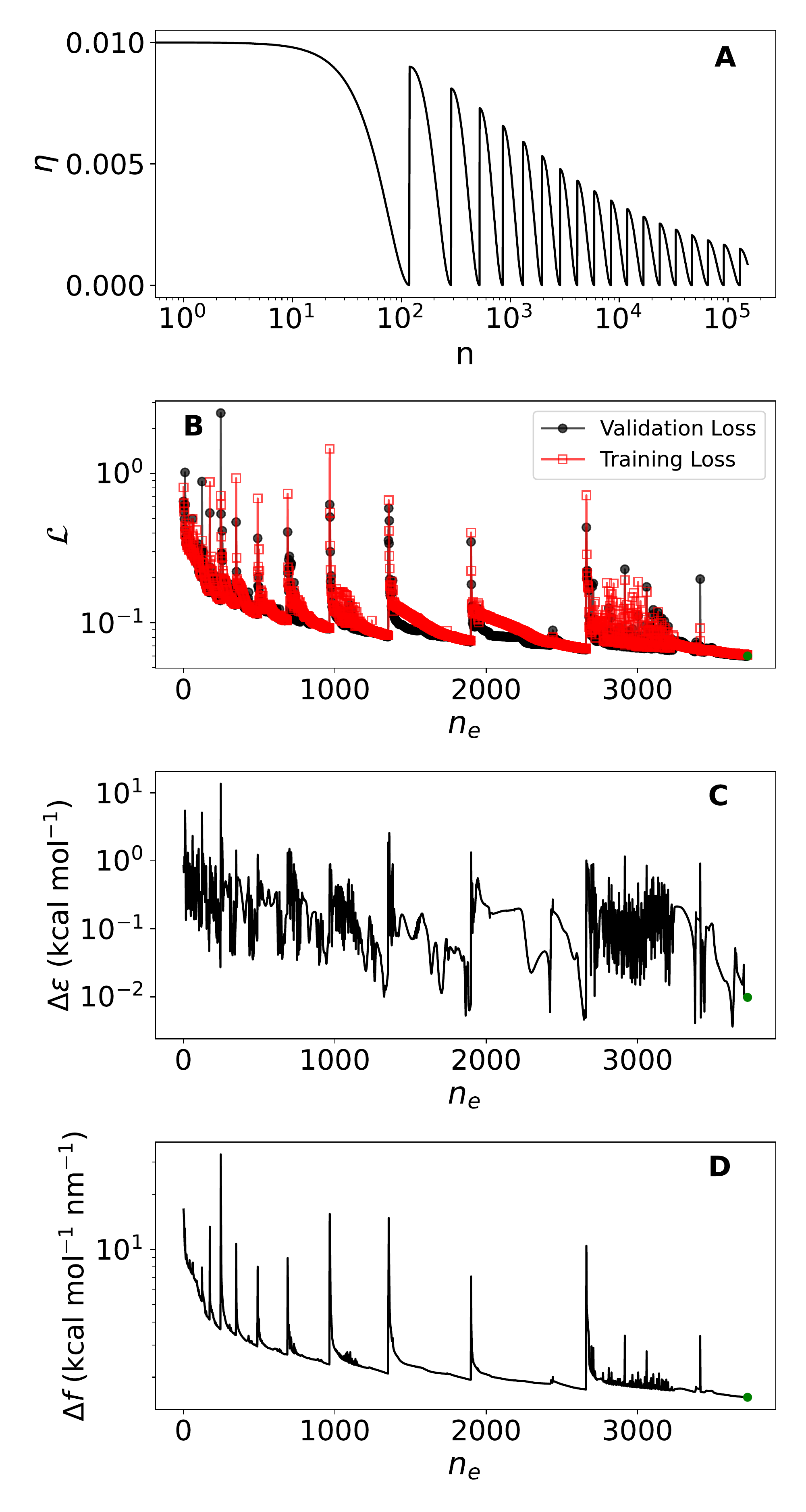} 
   \caption{Model convergence properties: (A) Learning rate  schedule (Eq.~\ref{eq:lr}) as a function of the absolute training  step $n$ (one step is defined as an update of the network parameters). (B) The training and validation loss (see $ \mathcal{L}[\epsilon,f] $ in Eq.~\ref{eq:huber}) evolution during the training procedure, reported as a function of the number of epoch $n_e$ (an epoch is defined as a complete evaluation of the training dataset). Root mean square (RMS) error of the total potential energy per particle (C) 
   and of the  force cartesian components (D) 
   during the training evaluated in the
   test dataset.   Data in panels B-C-D refers to the NN3 model and the green point shows the best model location.}
   \label{fig:fig_loss}
\end{figure}

 \begin{figure}[!t] %
   \centering
   \includegraphics[width=8.5cm]{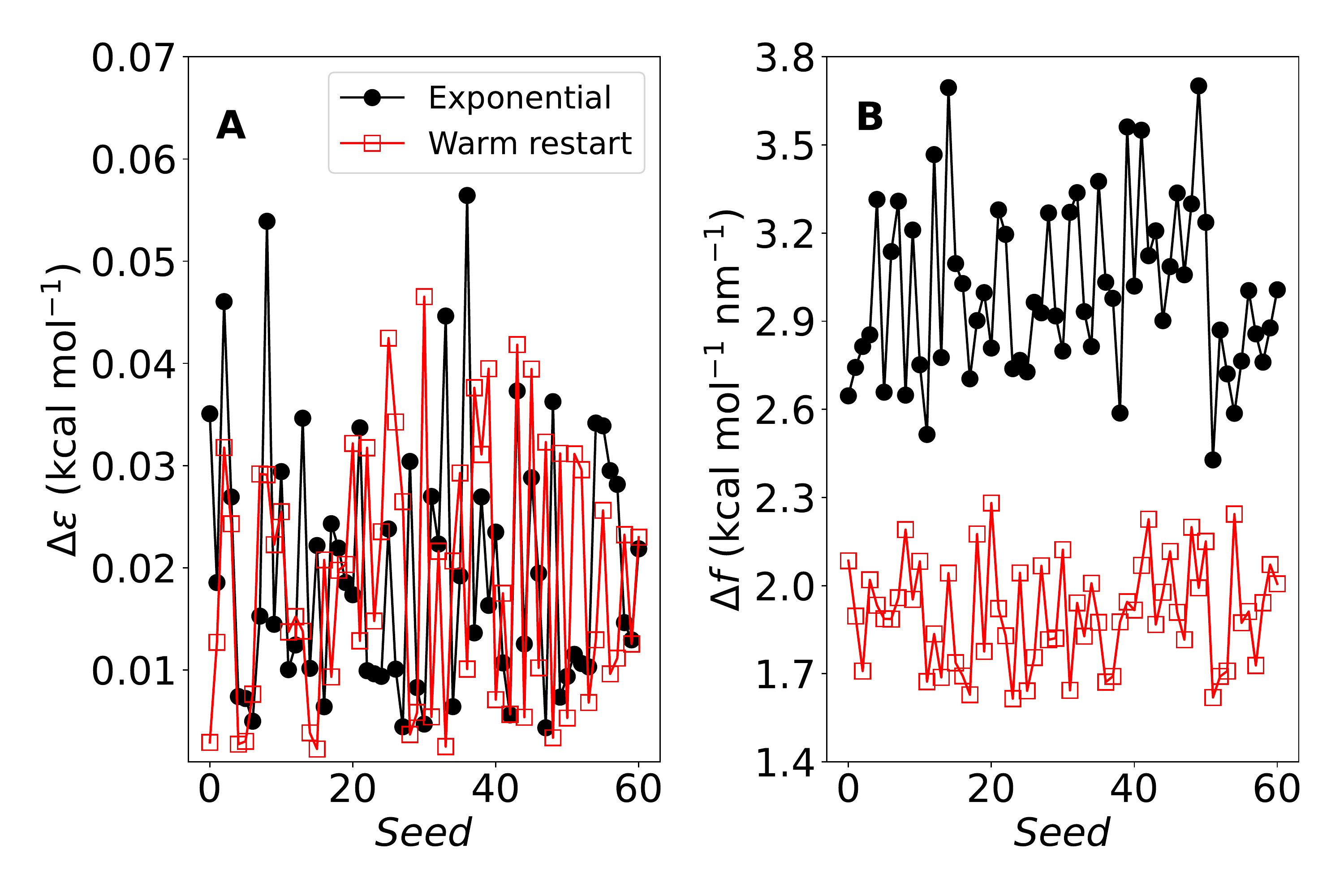} 
   \caption{Comparison of the root mean square error calculated on the validation set for 60 replicas differing in the initial seed of the training procedure using  both an exponential decay of the learning rate (points) and the warm restart method (squares), for the energy (panel A) and for the forces (panel B). For the forces, a significant improvement 
   both in the average error and in its variance is found for the warm restart schedule.}
   \label{fig:expann}
\end{figure}

In Fig.~\ref{fig:fig_loss}(A) we represent the typical decay of the learning rate of the warm restart procedure, which will be compared to the standard exponential decay protocol in the Results section.

\subsection{The Target Model}\label{sec:mW}
To test the quality of the proposed novel NN we train the NN with data produced with the 
mW~\cite{molinero2009water} model of water. This potential, a re-parametrization of the Stillinger-Weber model for silicon~\cite{stillinger1985computer}, uses a combination of pairwise functions complemented with an additive three-body potential term

\begin{equation} 
E=\sum_i\sum_{j>i}U_{2}(r_{ij})+\lambda\sum_{i}\sum_{j\neq i}\sum_{j>k}U_3\left(r_{ij},r_{ik},\theta_{jik}\right)
\end{equation} 

where the two body contribution between two particles $i$ and $j$ at relative distance $r_{ij}$ 
 is a generalized Lennard-Jones potential 

 \begin{equation}
U_2\left(r_{ij}\right)=A\epsilon\left[B\left(\frac{\sigma}{r_{ij}}\right)^p-\left(\frac{\sigma}{r_{ij}}\right)^q\right]exp\left(\frac{\sigma}{r_{ij}-a\sigma}\right)\end{equation} 

where the $p=12$ and $q=6$ powers are substituted by
$q=0$ and $p=4$,  multiplied by an exponential cut-off that brings the potential to zero at $a \sigma$, with $a=1.8$ and $\sigma=2.3925$~\AA. $A\epsilon$ (with $A=7.049556277$ and $\epsilon=6.189$~kcal~mol$^{-1}$) controls the  strength of the two body part. B controls the two-body repulsion (with $B= 0.6022245584$).

The three body contribution is computed from all possible ordered triplets formed by the central particle with the
interacting neighbors (with the same cut-off $a\sigma$ as the two-body term) and favours the
tetrahedral coordination of the atoms via the following functional form

 \begin{eqnarray}
U_3\left(r_{ij},r_{ik},\theta_{jik}\right)=\epsilon \left[\cos\left(\theta_{jik}\right)-\cos\left(\theta_{0}\right)\right]^2 \times  \nonumber \\
\exp\left(\frac{\gamma\sigma}{r_{ij}-a\sigma}\right)\exp\left(\frac{\gamma\sigma}{r_{ik}-a\sigma}\right)\label{eq:mw}
\end{eqnarray}

where $\theta_{jik}$ is the angle formed in the triplet $jik$ and $\gamma=1.2$  controls the smoothness of the cut-off function on approaching the cut-off. Finally, $\theta_{0}=109.47^{\circ}$ and $\lambda=23.15$ controls the strength of the 
angular part of the potential. 

The mW model, with its three-body terms centered around a specific angle and 
non-monotonic radial interactions, is based on a functional form which is quite different from the
radial and angular descriptors selected in the NN model.  The NN is thus agnostic with respect to the 
functional form that describes the physical system (the mW in this case). 
But having a reference model with explicit three body contributions offers a more challenging target for the NN potential compared to potential models built entirely from pairwise interactions. The mW  model is thus an excellent candidate to test the
performance of the proposed NN potential.

\section{RESULTS}\label{sec:results}

\subsection{Training}\label{sec:training}

We study two different NN models, indicated with the labels NN1 and NN3, differing in the 
number of state points included in the training set. These two models are built with a cut-off of $R_c=4.545$~\AA~ for the two-body atomic descriptors and a cut-off of $R_c'=4.306$~\AA~ for the three-body atomic descriptors. $R_c'$ is the same as the mW cutoff while $R_c$ was made slightly larger to mitigate the suppression of information at the boundaries by the cutoff functions. The NN1 model uses only training information based 
on mW equilibrium configurations from  one state point 
at $\rho_1=1.07$~g~cm$^{-3}$, $T_1=270.9$~K where the stable phase is the liquid.
The NN3 model uses  training information based on mW liquid configurations in three different state points, two state points at $\rho_1=0.92$~g~cm$^{-3}$, $T_1=221.1$~K and $\rho_2=0.92$~g~cm$^{-3}$, $T_2=270.9$~K  where the stable solid phase is the clathrate Si34/Si136~\cite{romano2014novel} and one state point at $\rho_3=1.15$~g~cm$^{-3}$, $T_2=270.9$~K. 

This choice of points in the phase diagram is aimed to improve agreement with the low temperature-low density as well as high density regions of the phase diagram. Importantly, all configurations come from either stable or metastable liquid state configurations. Indeed, the point at $\rho_2=0.92$~g~cm$^{-3}$, $T_2=270.9$~K is quite close to the limit of stability (respect to cavitation) of the liquid state.

To generate the training set, we simulate a system of $N=1000$ mW particles with a standard  molecular dynamics code in the NVT ensemble, where we use a time step of $4$~fs  and  run $10^7$ steps for each state point. From these trajectory, we randomly select $200$ configurations (frames) to create a dataset of positions, total energies and forces. We then split the dataset in the \emph{training} and in the \emph{test} data sets, the first one containing $80\%$ of the data. We then run the training for $4000$ epochs with a minibatch of $4$ frames. At the end of every epoch, we check if the validation loss is improved and  we save the model parameters.
In Fig.~\ref{fig:fig_loss} we plot the 
loss function for the training and test datasets (B), the root mean square error of the total energy per particle (C), and of the force (D) for the NN3 model. The results show that the learning rate schedule of Eq.~\ref{eq:lr} is very effective in reducing both the loss and error functions. 

Interestingly, the neural network seems to avoid overfitting (i.e. the validation loss is decreasing at the same rate as the loss on the training data), and the best model (deepest local minimum explored), in a given window of training steps, is always found at the end of that window, which also indicates that the accuracy could be further improved by running more training steps. Indeed we found that by increasing the number of training steps by one order of magnitude the error in the forces decreases by a further $30\%$. Similar accuracy of the training stage is obtained also for the NN1 model (not shown).

The training procedure always terminates with an error on the test set equal or less than $\Delta \epsilon\simeq 0.01$~kcal~mol$^{-1}$ ($0.43$~meV) for the energy, and of $\Delta f\simeq 1.55$~kcal~mol$^{-1}$~nm$^{-1}$\,($6.72$~meV~\AA$^{-1}$) for the forces. These values are comparable to the state-of-the-art NN potentials \cite{zhang2018end,cheng2019ab,gartner2020signatures,zhai2022short}, and within the typical accuracy of DFT calculations ~\cite{gillan2016perspective}.

We can compare the precision of our model with that of alternative NN potentials trained on a range of water models.
An alternative mW neural network potential has been trained on a dataset made of 1991 configurations of 128 particles system at different pressure and temperature (including both liquid and ice structures) with Behler-Parinello symmetry functions~\cite{singraber2019library}. The training of this model (which uses more atomic fingerprints and a larger cutoff radius) converged to an error in energy of  $\Delta \epsilon\simeq 0.0062$~kcal~mol$^{-1}$\,($0.27$\,meV), and $\Delta f\simeq 3.46$~kcal~mol$^{-1}$~nm$^{-1}$\,($15.70$\,meV~\AA$^{-1}$) for the forces.
In a recent work searching for liquid-liquid transition signatures in an ab-initio water NN model \cite{gartner2020signatures}, a dataset of configurations  spanning  a temperature range of $0-600$~K and  a pressure range of $0-50$~GPa was selected. For a system of 192 particles, the training converged to an error in energy of  $\Delta \epsilon\simeq 0.010$~kcal~mol$^{-1}$\,($0.46$\,meV), and $\Delta f\simeq 9.96$~kcal~mol$^{-1}$~nm$^{-1}$\,($43.2$\,meV~\AA$^{-1}$) for the forces. In the  NN model of MB-POL~\cite{zhai2022short}, a dataset spanning a temperature range from $198$~K to $368$~K  at ambient pressure was selected. In this case, for a system  of 256 water molecule, an accuracy of $\Delta\epsilon\simeq 0.01$~kcal~mol$^{-1}$\,($0.43$~meV) and $\Delta f\simeq 10$~kcal~mol$^{-1}$~nm$^{-1}$\,($43.36$~meV~\AA$^{-1}$) was reached.   Finally, the NN for water at $T=~300$~K used in Ref.~\cite{cheng2019ab}, reached precisions of $\Delta\epsilon\simeq 0.046$~kcal~mol$^{-1}$\,($2$~meV) and $\Delta f\simeq 25.36$~kcal~mol$^{-1}$~nm$^{-1}$\,($110$~meV~\AA$^{-1}$).

While a direct comparison between NN potentials trained on different reference potentials is not a valid test to rank the respective accuracies,
the comparisons above show that our NN potential reaches a similar precision in energies, and possibly an improved error in the force estimation.

The accuracy of the NN potential could be further improved by extending the size of the dataset and the choice of the state points.
In fact, while the datasets in Ref.~\cite{gartner2020signatures,zhai2022short,cheng2019ab}
have been built with optimized procedures,  the dataset used in this study was prepared by sampling just one (NN1) or three (NN3) state-points.
Also the size of the datasets used in the present work is smaller or comparable to the ones of   Ref.~\cite{gartner2020signatures,zhai2022short,cheng2019ab}.

In Fig.~\ref{fig:expann} we compare the error in the energies (A) and the forces (B) between sixty independent training runs using the standard exponential decay of the learning rate (points) and the warm restart protocol (squares). The figure shows that while the errors in the energy computations are comparable between the two methods, the warm restart protocol allows the forces to be computed with higher accuracy.  Moreover we found that the warm restart procedure is less dependent on the initial seed and that it reaches deeper basins than the standard exponential cooling rate.

\subsection{Comparing NN1 with NN3}\label{sec:NN1_NN3}

The NN potential model was implemented in a custom MD code that makes use of the tensorflow C API~\cite{capi}. We adopted the same time step ($4$ fs), the same number of particles ($N=1000$) and the same number of steps ($10^7$) as for the simulations in the mW model.

\begin{figure}[!t] %
   \centering
   \includegraphics[width=9cm]{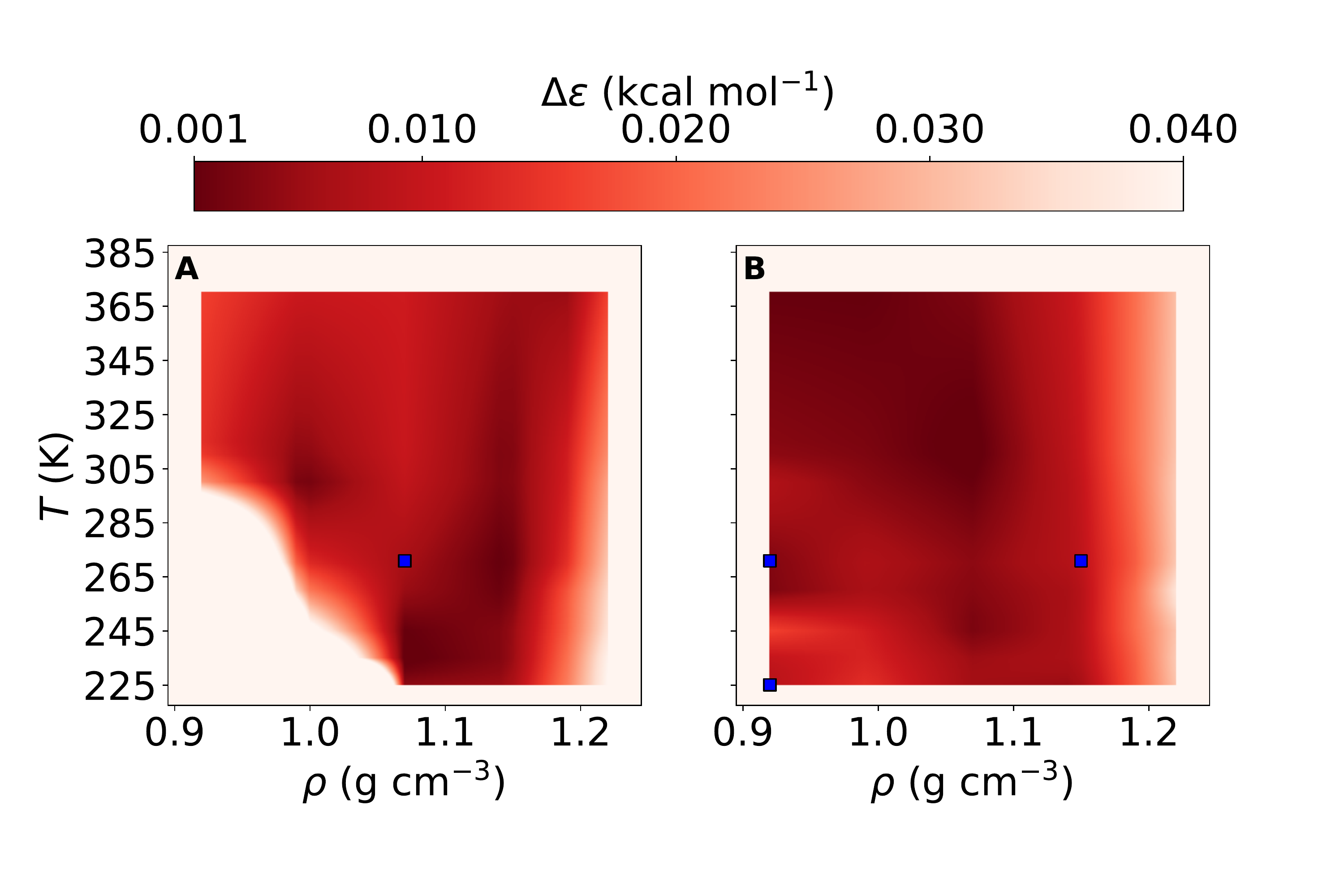} 
   \caption{Comparison between the mW total energy and the NN1 model (A) and NN3 model (B) for different  temperatures and densities.  While the NN3 model is able to reproduce the mW total energy with a good agreement in a wide region of densities and temperatures, the  NN1 provide a good representation only in a limited region of density and temperature values. Blue squares represent the state points used for building the NN models.}
   \label{fig:fig_stato}
\end{figure}

As described in the Training Section,
we compare the accuracy of two different training strategies: NN1 which was trained on a single state point, and NN3 which is instead trained on three different state point.
In Fig.~\ref{fig:fig_stato} we plot the energy error ($\Delta\epsilon$) between the NN potential and the mW model with both NN1 (panel A) and NN3 (panel B).
Starting from NN1, we see that the model already provides an excellent accuracy for a large range of temperatures and for densities close to the training density. The biggest shortcoming of the NN1 model is at densities lower than the trained density, where the NN potential model cavitates and does not retain the long-lived metastable liquid state displayed by the mW model. We speculate that this behaviour is due to the absence of low density configurations in the training set, which prevents the NN potential model from correctly reproducing the attractive tails of the mW potential.

To overcome this limitation we have included two additional state points at low density in the NN3 model. In this case, Fig.~\ref{fig:fig_stato}B shows that NN3
provides a quite accurate reproduction of the energy in the entire explored density and temperature window (despite being trained only with data at $\rho=0.92$~g~cm$^{-3}$ and $\rho=1.15$~g~cm$^{-3}$).

We can also compare the accuracy obtained during production runs against the accuracy reached during training, which was $\Delta\epsilon\simeq 0.01$~kcal~mol$^{-1}$.
Fig.~\ref{fig:fig_stato}B shows the error is of the order of $0.032$~kcal~mol$^{-1}$~($1.3$~meV), for density above the training set density. But in the density region between 0.92 and 1.15, the error is even smaller, around $0.017$~kcal~mol$^{-1}$~~($0.7$~meV) at the lowest density boundary.
 
We can thus conclude that the NN3 model, which adds to the NN1 model information at lower density and temperature, in the region where tetrahedality in the water structure is enhanced, is indeed capable to represent, with only three state points, a quite large
 region of the phase space, encompassing dense and stretched liquid states.  This suggests that 
 a training based on few state points at the boundary of the density/temperature region which needs to be studied is sufficient  to produce a high quality NN model.  In the following we focus entirely on the NN3 model.

 \subsection{Comparison of thermodynamic, structural and dynamical quantities}

\begin{figure}[!t] %
   \centering
   \includegraphics[width=9cm]{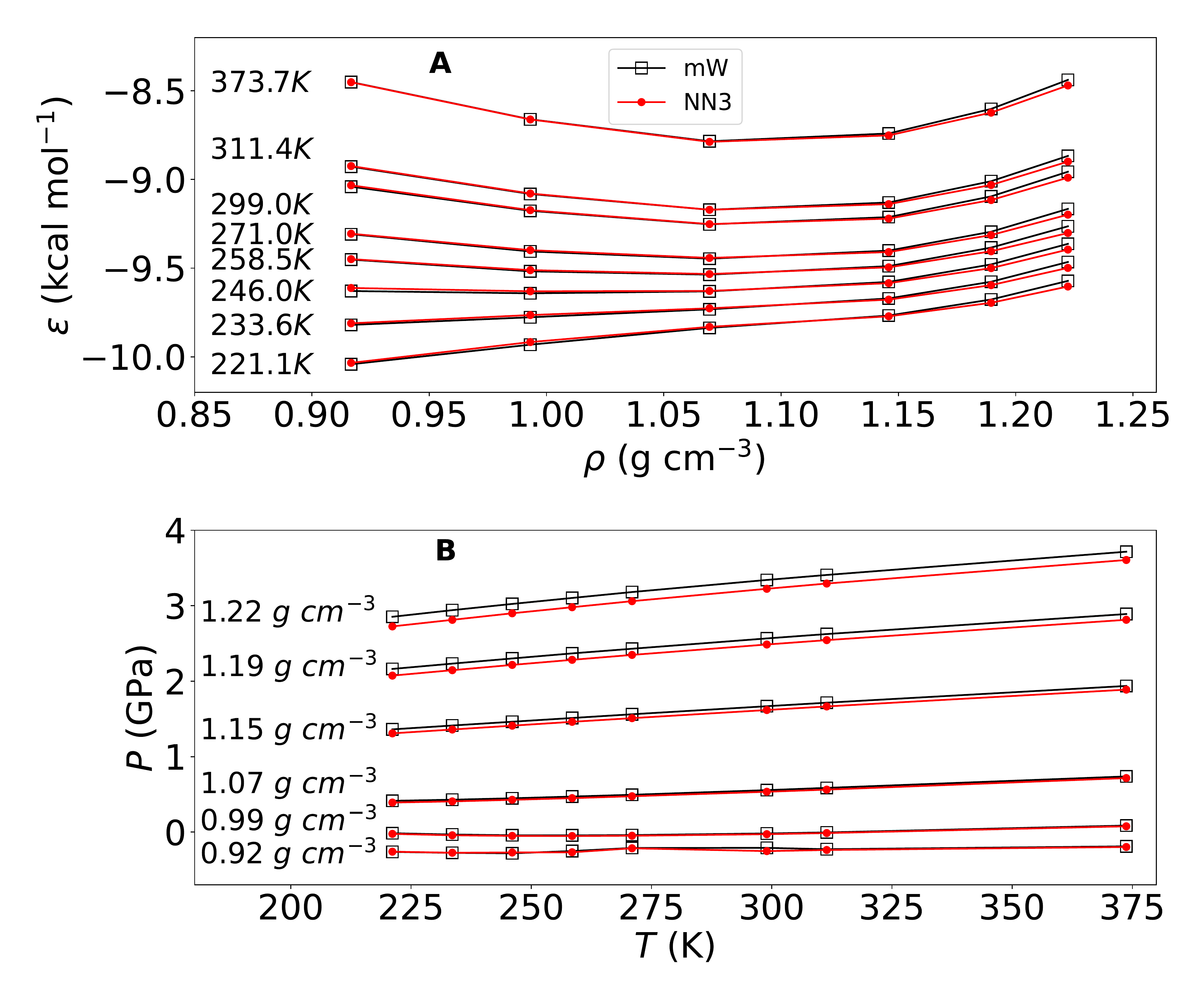} 
   \caption{Comparison between the mW  total energy and the NN3 total energy as a function of density along different isotherm (A) and comparison between the mW  pressure and the NN3 pressure as a function of temperature along different isochores (B). The relative error of the NN vs the mW potential grows with density, but remains within $3\%$ even for densities larger than the densities used in the training set.}
   \label{fig:pevst}
\end{figure}

In Fig.~\ref{fig:pevst} we present a comparison of thermodynamic data between the mW model (squares) and its NN potential representation (points) across a wide range of state points.
Fig.~~\ref{fig:pevst}A plots the energy as function of density for temperatures ranging from melting to deeply supercooled conditions. Perhaps the most interesting result is that the NN potential is able to capture the energy minimum, also called the \emph{optimal network forming density}, which is a distinctive anomalous property of water and other \emph{empty liquids}~\cite{russo2021physics}.
 
 Fig.~\ref{fig:pevst}(B) shows the pressure as a function of the temperature for different densities, comparing the mW with the NN3 model. Also the pressure shows a good agreement between the two models in the region of densities between $\rho=0.92$~g~cm$^{-3}$ and $\rho=1.15$~g~cm$^{-3}$, which, as for the energy, tends to deteriorate at $\rho=1.22$~g~cm$^{-3}$.

\begin{figure}[!t] %
   \centering
   \includegraphics[width=9cm]{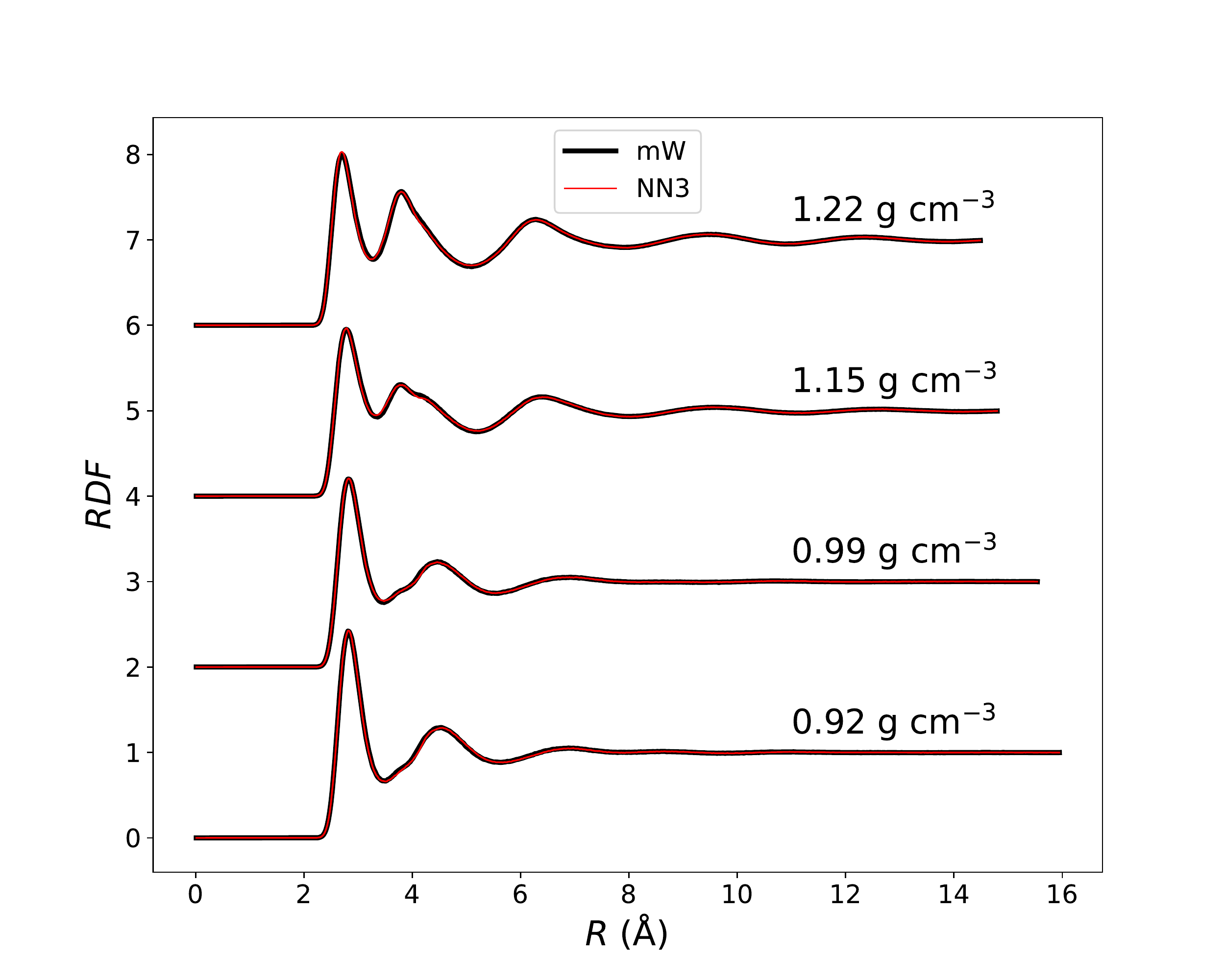} 
   \caption{Comparison  between the mW  radial distribution functions $g(r)$ and the NN3 $g(r)$  at $T =270.9$ K for four different densities. The tetrahedral structure (signalled by the peak at $4.54$~\AA~) progressively weakens in favour of an interstitial peak progressively growing at $3.5-3.8$~\AA~. Different $g(r)$ have been progressively shifted by two to improve clarity.}
   \label{fig:fig_gr}
\end{figure}

 In the large density region explored, the structure of the liquid changes considerably. On increasing density, a transition from tetrahedral coordinated  local structure, prevalent at low $T$ and low $\rho$, towards denser local environments with interstitial molecules included in the first coordination shell takes place. This structural change is well displayed in the 
 radial distribution function, shown for different densities at fixed temperature in Fig.~\ref{fig:fig_gr}.   
 Fig.~\ref{fig:fig_gr} also shows the progressive onset of a peak around $3.5$~\AA~ 
 developing on increasing pressure, which signals the growth of interstitial molecules, coexisting with open tetrahedral local structures~\cite{foffi2021structure,foffi2021structural}.  At the highest density, the tetrahedral peak completely merges with the interstitial peak.
The NN3 model reproduces quite accurately all features of the radial distribution functions, maxima and minima positions and their relative amplitudes, at all densities, from the tetrahedral-dominated  to the interstitial-dominated limits.
In general, NN3 model reproduces quite well the mW potential in energies, pressure and structures and it appreciably deviates from mW pressures and energies quantities only at densities (above 1.15 g/cm$^3$) which are outside of the training region.

\begin{figure}[!t] %
   \centering
\includegraphics[width=8.5cm]{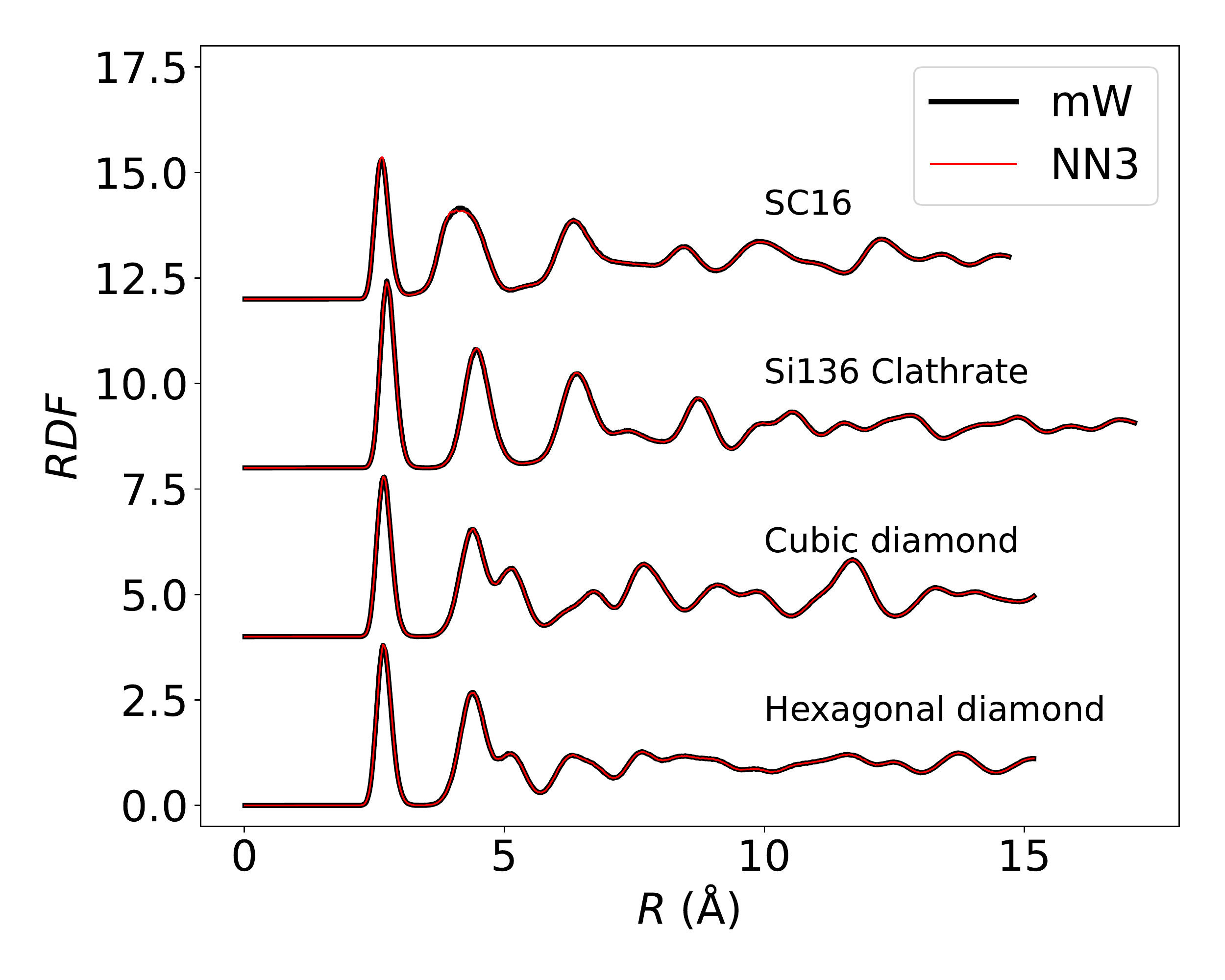} 
   \caption{Comparison  between the mW  radial distribution functions $g(r)$ and the NN3 $g(r)$  for four different lattices:
   (A) hexagonal diamond (the oxygen positions of the ice $I_h$); (B) cubic diamond (the oxygen positions of the ice $I_c$; (C) the SC16 crystal (the dense crystal form stable at large pressures in the mW model) and (D) the Si136 clathrate structure, which is stable at negative pressures in the mW model. Different $g(r)$ have been progressively shifted by four to improve clarity.}
   \label{fig:fig_grxt}
\end{figure}

To assess the ability of NN potential to correctly describe also the crystal phases of the mW potential, we 
compare in Fig.~\ref{fig:fig_grxt} the $g(r)$ of mW with the $g(r)$ of the NN3 model for four different stable solid phases~\cite{romano2014novel}:
hexagonal and cubic ice  ($\rho=1.00$~g~cm$^{-3}$ and $T=246$~K), the dense crystal SC16  ($\rho=1.20$~g~cm$^{-3}$ and $T=234$~K) and  the clathrate phase Si136  ($\rho=0.80$~g~cm$^{-3}$ and $T=221$~K).  The results, shown in Fig.~\ref{fig:fig_grxt}, show that, despite no crystal configurations have been included in the training set, a quite accurate representation of the crystal structure at finite temperature is provided by the NN3 model for all distinct sampled lattices.

\begin{figure}[!t] %
   \centering
   \includegraphics[width=8.5cm]{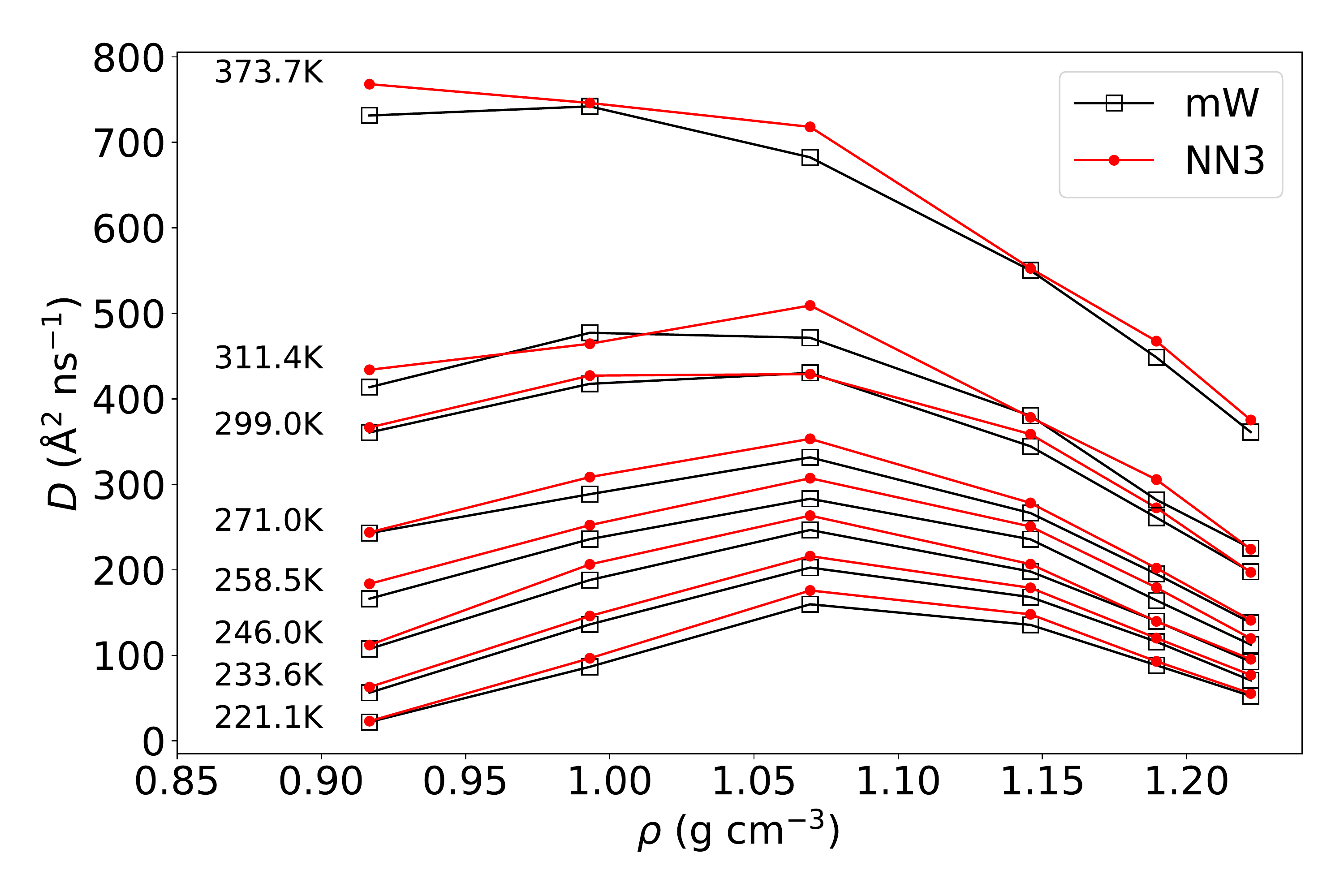} 
   \caption{Comparison  between the mW diffusion coefficient  $D$ and the NN3 corresponding quantity for different temperatures and densities, in the interval $221-271$ K.  In this dynamic quantity, the relative error is, for all temperatures, around $8\%$.  Note also that 
    in this $T$ window the diffusion coefficient shows a clear maximum, reproducing one of the well-know diffusion anomaly of water. Diffusion coefficients have been calculated in the NVT ensemble using the same Andersen thermostat algorithm~\cite{andersen1980molecular} for mW and NN3 potential.
    }
   \label{fig:diff}
\end{figure}

Finally, we compare in Fig.~\ref{fig:diff} the diffusion coefficient (evaluated from the long time limit of the mean square displacement) for the mW and the NN3 model, in a wide range of temperatures and densities, where water displays a diffusion anomaly. 
Fig.~\ref{fig:diff} shows again that, also for dynamical quantities, the NN potential offers an excellent representation of the mW potential, despite the fact that no dynamical quantity was included in the training set. A comparison between fluctuations of energy and pressure of mW and NN3 potential is reported in Appendix B.

\section{CONCLUSIONS}\label{sec:conclusions}

In this work we have presented a novel neural network (NN) potential based on a new set of atomic fingerprints (AFs) built from two- and three-body local descriptors that are combined in a permutation-invariant way through an exponential filter (see Eq.~\ref{eq:proj2b}-\ref{eq:proj3b}). One of the distinctive advantages of our scheme is that the AF's parameters are optimized during the training procedure, making the present algorithm a self-training network
that automatically selects the best AFs for the potential of interest.

We have shown that the added complexity in the concurrent training  of the AFs and of the NN weights can be overcome with an annealing procedure based on the warm restart method~\cite{loshchilov2016sgdr}, where the learning rate goes through damped oscillatory ramps. This strategy not only gives better accuracy compared to the commonly implemented exponential learning rate decay,
but also allows the training procedure to converge rapidly independently from the initialisation strategies of the model's parameters.

Moreover we show in Appendix C that the potential hyper-surface of the NN model has the same smoothness as the target model, as confirmed by  (i) the
possibility to use the same timestep in the NN and in the target model when integrating the equation of motion and (ii) by the possibility of simulate the NN model even in the NVE ensemble with proper energy conservation.

We test the novel NN on the mW model~\cite{molinero2009water}, a one-component model system commonly used to describe water in classical simulations.
This model, a re-parametrization of the Stillinger-Weber model for silicon~\cite{stillinger1985computer}, 
 while treating the water molecule as a simple point, is able to reproduce the characteristic tetrahedral local structure of water
(and its distortion on increasing density) via the use of three-body interactions. Indeed water changes from a liquid of tetrahedrally coordinated molecules to a denser liquid, in which a relevant fraction of interstitial molecules  are present in the first nearest-neighbour shell.
The complexity of the  mW model, both due to its functional form as well as to the variety of  different local structures which characterise water, makes it an ideal benchmark system to test our NN potential.

We find that a training based on configurations extracted by three different state points is able to provide a quite accurate representation of the mW potential hyper-surface, when the  densities  and temperatures of the training state points delimit the region of in which the NN potential is expected to work. 
We also find that the error in the NN estimate  of the total energy is low, 
always smaller than $0.03$~kcal~mol$^{-1}$, with a mean error of $0.013$~kcal~mol$^{-1}$. The NN model reproduces very well not only the thermodynamic properties but also structural properties, as quantified by the radial distribution function, and the dynamic properties, as expressed by the diffusion coefficient, in the extended density interval from
$\rho=0.92$~g~cm$^{-3}$ to $\rho=1.22$~g~cm$^{-3}$.

Interestingly, we find that the NN model, trained only on disordered configurations, is also able to properly describe the
radial distribution of  the  ordered lattices which characterise the mW phase diagram, encompassing the 
cubic and hexagonal ices, the SC16 and the Si136 clathrate structure~\cite{romano2014novel}. In this respect, the ability of the
NN model to properly represent crystal states suggests that, in the case of the mW, and as such probably in the case of water,
the geometrical information relevant to the ordered structures is contained in the sampling of phase space typical of the disordered liquid phase. These findings have been recently discussed in reference~\cite{monserrat2020liquid} where it has been demonstrated that liquid water contains all the building blocks of diverse ice phases.

We conclude by noticing that the present approach can be generalized to multicomponent systems, following the same
strategy implemented by previous approaches~\cite{behler2007generalized,zhang2018end}. Work in this direction is underway.

\begin{acknowledgments}
FGM and JR acknowledge support from the European Research Council Grant DLV-759187 and CINECA grant 
ISCRAB NNPROT.
\end{acknowledgments}


\providecommand{\noopsort}[1]{}\providecommand{\singleletter}[1]{#1}%

\section{APPENDIX A}\label{sec:appendix}

\setcounter{equation}{0}
\renewcommand{\theequation}{A\arabic{equation}}

In this appendix we discuss the effective spacial range covered by a NN potential whose fingerprints are defined 
based on pair information confined within a sphere of cutoff radius $R_c$.

As noted in reference \cite{behler2021four}, multi-body potentials and especially non-additive multibody potentials induce local interactions beyond the cut-off radius, enlarging the sphere of interaction. 
Indeed, the force on particle $i$ comes from the derivative of the local field of $i$ and of all its neighbours with respect to
the coordinates of  particle $i$. 

\begin{figure}[htbp] %
   \centering
   \includegraphics[width=8.5cm]{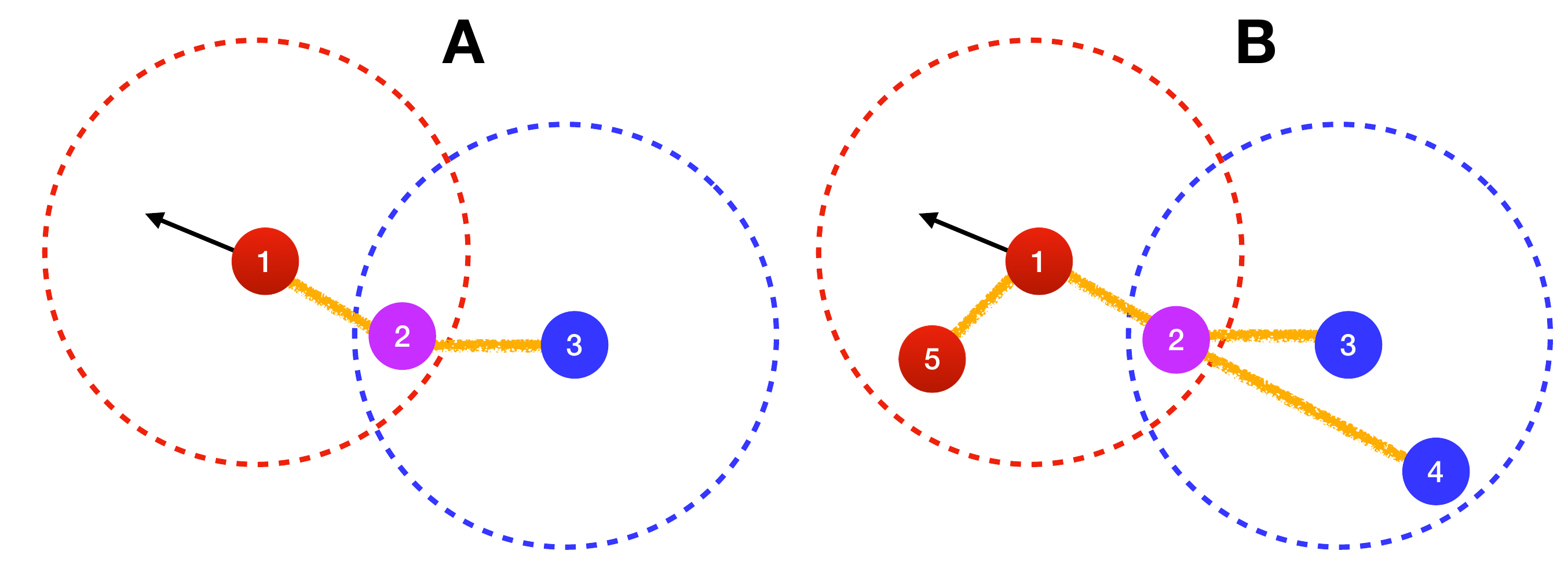} 
   \caption{(A) Two-body interactions and (B) three-body interactions in a non linear local field model $E_i$. The non linearity of the local field enlarges the interaction cut-off where a neighbour particle (blue) makes a bridge between non-neighboring particle (red and blue).}
   \label{fig:inter}
\end{figure}

Fig.~\ref{fig:inter} graphically explains the effective role of $R_c$ in the NN potential.  In panel A, 
we describe particle 1 with only one neighbour (particle 2) within $R_c$. We also represent the sphere centered on 
particle $3$, which also includes particle 2 as one of its neighbour.
In this case, the energy of the system will be represented as a sum over the local fields $E_1,E_2$ and $E_3$.
Due to the intrinsic non-linearity of the NN, the field $E_i$ mixes together the AFs, and consequently the distances and angles entering in the AFs are non-linearly mixed in $E_i$. The force on atom 1 is then written as

\begin{eqnarray}
f_{1\nu}=-\frac{\partial E_1(r_{12})}{\partial x_{1\nu}}-\frac{\partial E_2(r_{21},r_{23})}{\partial x_{1\nu}}=  
-\frac{\partial E_1(r_{12})}{\partial x_{1\nu}} \nonumber \\
-\frac{\partial E_2(r_{21},r_{23})} {\partial r_{21}} \frac{\partial r_{21}} {\partial x_{1\nu}} -
\frac{\partial E_2(r_{21},r_{23})} {\partial r_{23}} \frac{\partial r_{23}} {\partial x_{1\nu}}
\end{eqnarray}
While the last term vanishes, the next to the last retains an intrinsic dependence on the coordinates both of particle 2 as well as of particle 3, if the local field $E_2$ is non linear.
Thus, even if particle 3 is further than $R_c$, it enters in the determination of the force acting on particle 1.
A similar effect is also present in the angular part of the AFs, as shown graphically in panel B.
Indeed, for  the angular component of the  AF  the force on particle 1 is
\begin{equation}
f_{1\nu}=-\frac{\partial E_1(\theta_{512})}{\partial x_{1\nu}}-\frac{\partial E_2(\theta_{123},\theta_{124},\theta_{324})}{\partial x_{1\nu}}.
\end{equation}
Also in this case two contributions can be separated: (i) the interaction of particle 1 with triplets $123$ and $124$ is an effect of the three-body AF and it is present also in additive-models such as the mW model, (ii) the interaction of particle 1 with triplet $324$ is an effect of the non-additive nature of the NN local field $E_i$.  

\section{APPENDIX B}\label{sec:appendix}

\renewcommand{\theequation}{B\arabic{equation}}

\begin{figure}[t!] %
   \centering
   \includegraphics[width=8.5cm]{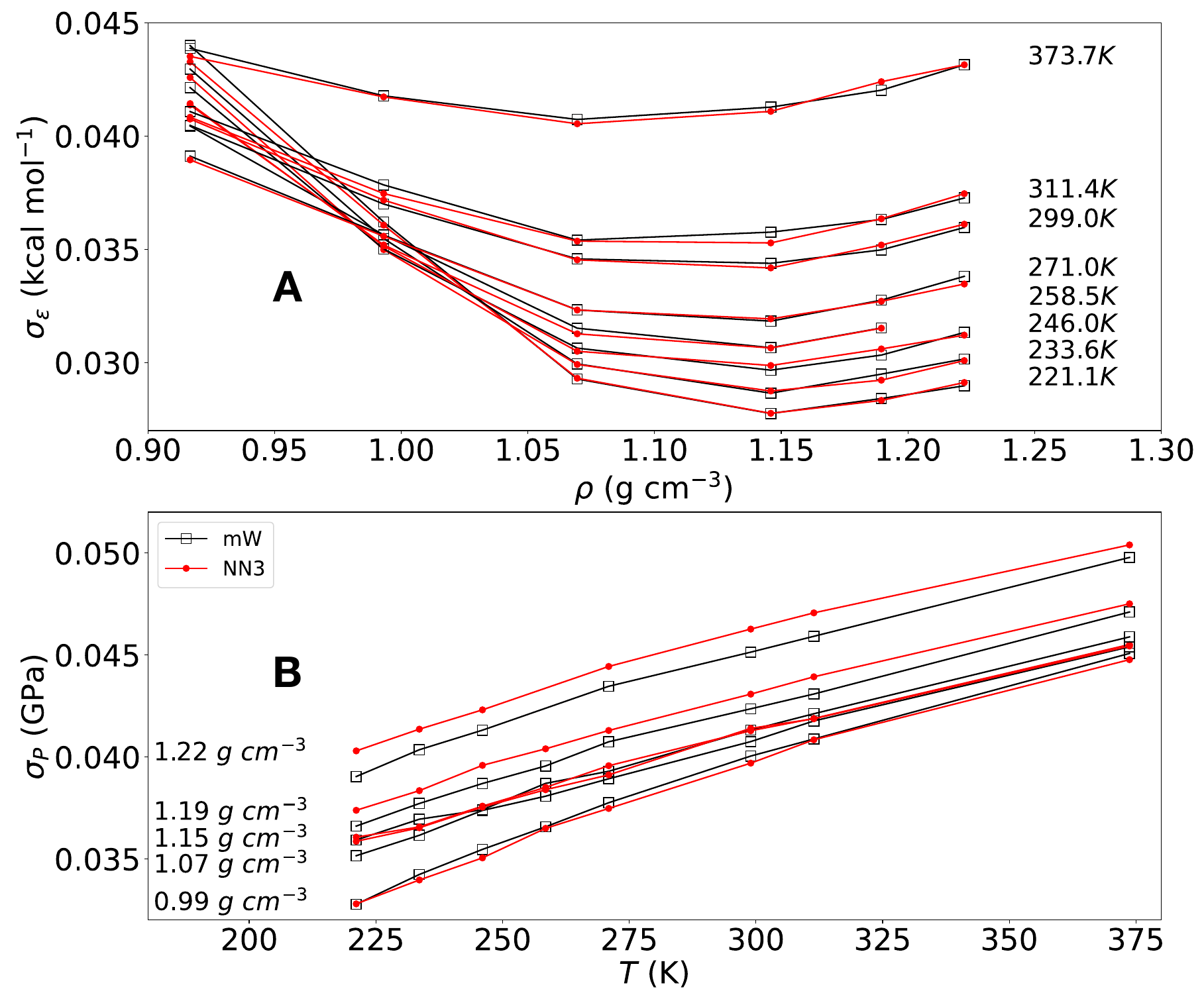} 
   \caption{(A) Standard deviation of total energy (normalized with the number of particles) and (B) standard deviation of virial pressure for both NN3 model (red) and mW model (black).}
   \label{fig:fluct}
\end{figure}

In this Appendix we provide further thermodynamics comparisons between mW and NN3 potential focusing on the pressure and energy fluctuations. We depict in Fig.~\ref{fig:fluct} the standard deviations of the total energy (normalized by N) in panel (A) and the standard deviation of virial pressure in panel (B). Energy fluctuations of NN3 follow qualitatively and quantitatively the trend of mW potential. Pressure fluctuations of NN3 are in good agreement with the mW model but, as for the pressure (Fig.~\ref{fig:pevst}.B), the accuracy decreases approaching state points outside the density range used for the training.

\section{APPENDIX C}\label{sec:appendix}

\renewcommand{\theequation}{C\arabic{equation}}

\begin{figure}[!t] %
   \centering
   \includegraphics[width=8.5cm]{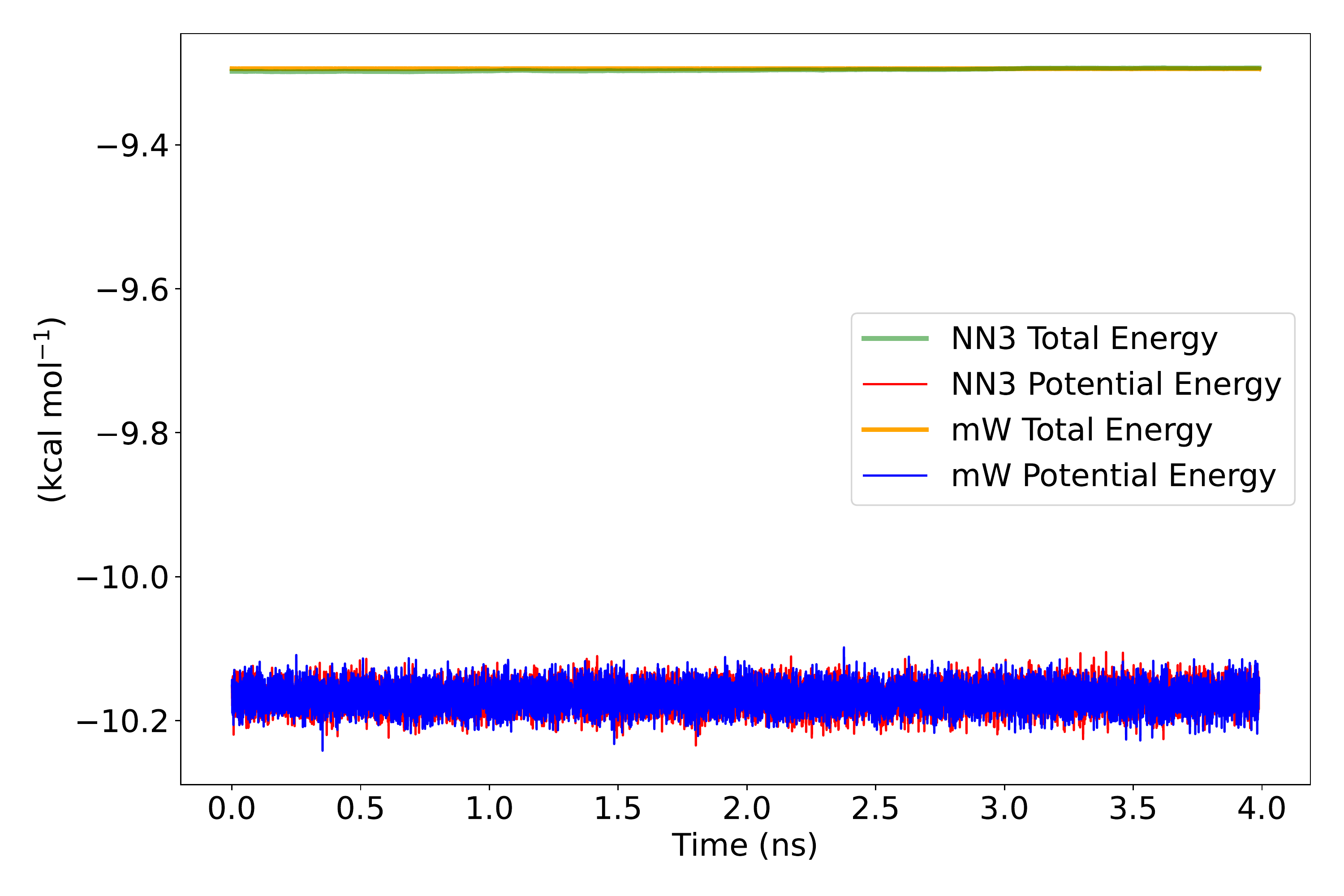} 
   \caption{NVE molecular dynamics at $T=299$~K and $\rho=1.07$~g~cm$^{-3}$ for both NN3 and mW model. The time step is $dt=4$~fs for both models.}
   \label{fig:nve}
\end{figure}

In this Appendix we show a comparison between the mW and NN3 potentials in terms of the energy conservation in the NVE ensemble. In Fig.~\ref{fig:nve} we depict both total energy and potential energy for mW and NN3 potential. The potential energy and total energy of the two models are in good agreement.

\section{APPENDIX D}\label{sec:appendix}

\renewcommand{\theequation}{D\arabic{equation}}

In this Appendix we investigate the efficiency of the training over different choices for the number and types of atomic fingerprints introduced in the Neural Network Model section.
We start by using only one three-body ($n_{3b}=1$) and one two-body ($n_{2b}=1$) AF and subsequently increasing the number of the AF. For every combination of $n_{2b}$ and $n_{3b}$, we run a 4000 epochs training and at the end of each training we extract the best model. We summarized these results in table \ref{tab:eff} where we compare the error on forces over the all investigated model. From table \ref{tab:eff} it emerges that the choice of $n_{3b}=5$ and $n_{2b}=5$ is the more convenient both for accuracy and computational efficiency. Doubling the number of the three-body AF marginally improves the error on forces while increases the computational cost due to the increase in the size of the input layer of the first hidden layer and due to the additional time to compute the three-body AF. Moreover in the RESULTS section we show that the choice $n_{3b}=5$ and $n_{2b}=5$ is sufficient to represent the target potential. Finally the accuracy of the training after doubling the configurations in the dataset reaches an error on forces of $\Delta_f=5.85$~meV~\AA$^{-1}$ that is 0.87 times  the error value found with a half of the dataset.

\begin{table}[b]
\caption{\label{tab:eff}
Table of errors on forces at the end of the 4000 epoch-long training procedure for different combination of the number and type 
of the AF.}
\begin{ruledtabular}
\begin{tabular}{cccccccc}
 &$n_{3b}$ &$n_{2b}$ &$\Delta f$~(meV~\AA$^{-1}$)&
 &$n_{3b}$ &$n_{2b}$ &$\Delta f$~(meV~\AA$^{-1}$)\\
\hline
& 1 & 1 & 72.79 &
& 5 & 1 & 16.53 \\
& 1 & 2 & 67.92 &
& 5 & 2 & 7.53 \\
& 1 & 5 & 56.25 &
& 5 & 5 & 6.72 \\
& 1 & 10 & 56.00 &
& 5 & 10 & 6.87 \\
& 1 & 15 & 56.02 &
& 5 & 15 & 6.95 \\
& 2 & 1 & 53.76 &
& 10 & 1 & 7.98\\
& 2 & 2 & 43.95 &
& 10 & 2 & 7.17 \\
& 2 & 5 & 32.43 &
& 10 & 5 & 5.79 \\
& 2 & 10 & 32.39 &
& 10 & 10 & 6.55\\
& 2 & 15 & 24.70 &
 & 10 & 15 & 6.19\\
\\
\end{tabular}
\end{ruledtabular}
\end{table}

\end{document}